\documentclass[12pt]{article}
\usepackage{epsfig,amssymb,amsmath,psfrag}

\catcode`\@=11
\newcommand{\insertfig}[2]{\mbox{\epsfxsize=#1cm \epsfbox{#2.eps}}}

\font\cmss=cmss12 
\def\1{\hbox{{1}\kern-.25em\hbox{l}}}
\def\bfZ{\relax{\hbox{\cmss Z\kern-.4em Z}}}

\def \be  {\begin{equation}}
\def \ee  {\end{equation}}
\def \ba  {\begin{eqnarray}}
\def \ea  {\end{eqnarray}}
\def \baa {\begin{eqnarray*}}
\def \eaa {\end{eqnarray*}}
\def \bb  {\begin {thebibliography} }
\def \eb  {\end{thebibliography}}
\def \lab #1 {\label{#1}}

\newcommand\re[1]{(\ref{#1})}

\def \qqqquad {\qquad\qquad}
\def \matrix #1 {\left(\begin{array}{cc} #1 \end{array}\right)}

\def \e  {\mathop{\rm e}\nolimits}
\newcommand\lr[1]{{\left({#1}\right)}}
\newcommand \widebar [1] {\overline{#1}}

\newcommand{\as}{\ifmmode\alpha_{\rm s}\else{$\alpha_{\rm s}$}\fi}
\newcommand{\asbar}{\ifmmode\bar{\alpha}_{\rm s}\else{$\bar{\alpha}_{\rm s}$}\fi}

\newcommand{\bit}[1]{\mbox{\boldmath$#1$}}
\newcommand{\ft}[2]{{\textstyle\frac{#1}{#2}}}

\font\cmss=cmss12 
\def\inbar{\,\vrule height1.5ex width.4pt depth0pt}
\def\IC{\relax\hbox{$\inbar\kern-.3em{\rm C}$}}
\def\IZ{\relax{\hbox{\cmss Z\kern-.4em Z}}}
\def\IR{{\hbox{{\rm I}\kern-.2em\hbox{\rm R}}}}

\def\IP{{\hbox{{\rm I}\kern-.2em\hbox{\rm P}}}}
\def\II{\hbox{{1}\kern-.25em\hbox{l}}}

\def\numberbysection{\@addtoreset{equation}{section}
                     \def\theequation{\thesection.\arabic{equation}}}
\numberbysection

\newbox\lett\newdimen\lheight\newdimen\lwidth
\def\ontop#1#2{\setbox\lett=\hbox{#2}\lheight\ht\lett
\multiply\lheight by 12 \divide\lheight by 10\relax%
\lwidth\wd\lett \multiply\lwidth by 8 \divide\lwidth by 10\relax #2\kern-\lwidth%
\raise\lheight\hbox{{$\scriptstyle #1$}}\kern.1ex}

\def\XXint#1#2#3{{\setbox0=\hbox{$#1{#2#3}{\int}$}
     \vcenter{\hbox{$#2#3$}}\kern-.5\wd0}}

\psfrag{dots}[cc][cc]{$...$}
\psfrag{Z1}[cc][cc]{$\scriptstyle Z_1$}
\psfrag{Z2}[cc][cc]{$\scriptstyle Z_2$}
\psfrag{Z3}[cc][cc]{$ $}
\psfrag{ZN-1}[cc][cc]{$ $}
\psfrag{ZN}[cc][cc]{$\scriptstyle Z_L$}

\begin{document}

\begin{titlepage}
\begin{flushright}
\begin{tabular}{l}
LPT--Orsay--06--36
\end{tabular}
\end{flushright}

\vskip1cm

\centerline{\large \bf Towards Baxter equation in supersymmetric Yang-Mills theories}

\vspace{1cm}

\centerline{\sc A.V. Belitsky$^a$, G.P. Korchemsky$^b$, D. M\"uller$^a$}

\vspace{10mm}

\centerline{\it $^a$Department of Physics and Astronomy, Arizona State University}
\centerline{\it Tempe, AZ 85287-1504, USA}

\vspace{3mm}

\centerline{\it $^b$Laboratoire de Physique Th\'eorique\footnote{Unit\'e
                    Mixte de Recherche du CNRS (UMR 8627).},
                    Universit\'e de Paris XI}
\centerline{\it 91405 Orsay C\'edex, France}

\def\thefootnote{\fnsymbol{footnote}}%
\vspace{1cm}

\centerline{\bf Abstract}

\vspace{5mm}

We perform an explicit two-loop calculation of the dilatation operator acting on
single trace Wilson operators built from holomorphic scalar fields and an arbitrary
number of covariant derivatives in $\mathcal{\mathcal{N}}=2$ and $\mathcal{N}=4$
supersymmetric Yang-Mills theories. We demonstrate that its eigenspectrum exhibits
double degeneracy of opposite parity eigenstates which suggests that the two-loop
dilatation operator is integrable. Moreover, the two-loop anomalous dimensions in
the two theories differ from each other by an overall normalization factor
indicating that the phenomenon is not sensitive to the presence of the conformal
symmetry. Relying on these findings, we try to uncover integrable structures behind
the two-loop dilatation operator using the method of the Baxter $\mathbb{Q}-$operator.
We propose a deformed Baxter equation which exactly encodes the spectrum of two-loop
anomalous dimensions and argue that it correctly incorporates a peculiar feature of
conformal scalar operators -- the conformal $SL(2)$ spin of such operators is modified
in higher loops by an amount proportional to their anomalous dimension.

\end{titlepage}

\setcounter{footnote} 0

\thispagestyle{empty}

\newpage

\pagestyle{plain} \setcounter{page} 1

\section{Introduction}

The analysis of one-loop anomalous dimensions of composite high-twist operators in QCD
revealed that its dilatation operator possesses nontrivial integrability symmetry. It
was observed in Refs.\ \cite{BraDerMan98,Bel99,DerKorMan00} that the one-loop mixing matrix
for the so-called maximal-helicity quasipartonic \cite{BukFroKurLip85} Wilson operators
can be mapped in the multi-color limit into the Hamiltonian of the $SL(2;\mathbb{R})$
Heisenberg spin chain and its eigenspectrum can be computed exactly with the help of
Bethe Ansatz. The length of the chain is determined by the number of elementary fields
in the Wilson operator and the spin operators on its sites are defined by the generators
of the collinear subgroup of the conformal group $SO(4,2)$. Let us emphasize that similar
structures were discovered earlier in the Regge limit of QCD \cite{Lip93,FadKor95}.
Integrability imposes a very nontrivial analytic structure on the anomalous dimensions,
which is reflected, in particular, in pairing of opposite-parity eigenstates in the spectrum.

Integrability observed in QCD anomalous dimensions at one-loop order is a generic
phenomenon of four-dimensional Yang-Mills theories and it is ultimately related to the
presence of massless spin-one gauge bosons in the particle spectrum. It is also present,
as was found in Refs.\ \cite{Lip98,MinZar02,BeiSta03}, in maximally supersymmetric
Yang-Mills (SYM) theory. Theories with less supersymmetries also inherit integrability
although the number of integrable sectors strongly depends on the particle content of the
models and is enhanced for theories with more supercharges \cite{BelDerKorMan04}. In this
regard the maximally supersymmetric $\mathcal{N} = 4$ Yang-Mills theory currently occupies
a distinguished niche in light of the gauge/string duality \cite{Mal97} which allows one
to establish the correspondence between the anomalous dimensions of composite operators in
$\mathcal{N}=4$ theory and energies of string excitations on the AdS${}_5 \times$S${}^5$
background \cite{Pol01,BerMalNas02,GubKlePol03}. It was recently shown that classical string
sigma models with anti-de Sitter space as a factor of the target space possess an infinite
set of integrals of motion and therefore are integrable \cite{ManSurWad02,BeRoiPol03,Pol04}.
On gauge theory side, this suggests that the dilatation operator for Wilson operators carrying
large quantum numbers should be integrable in the $\mathcal{N}=4$ theory in the strong
coupling regime.

The question arises whether the one-loop integrability of the dilatation operator and
integrability of the classical string sigma model is a manifestation of the same universal
phenomenon at weak and strong coupling, respectively, and if so then whether the
``perturbative'' dilatation operator exhibits integrability order-by-order in the coupling
constant. Since the range of interaction in the spin chain increases with order in 't Hooft
coupling constant $\lambda=g_{\rm \scriptscriptstyle YM}^2 N_c/(8 \pi^2)$, -- being merely
nearest-neighbor at one loop, then stretching to three adjacent neighbors at two loops, etc.
-- the spectrum of anomalous dimensions should be determined by yet unidentified long-range
integrable spin chain. Recent extensive perturbative studies indeed support this conjecture
\cite{BeiKriSta03,Bei04,EdeJarSok04,Sta04,Ede04,BelKorMul04,EdeJarSokSta04,Zwi06,EdeSta06}.

In this paper, we perform an explicit two-loop calculation of the dilatation operator acting
on single-trace Wilson operators built from holomorphic scalar fields and an arbitrary number
of covariant derivatives in $\mathcal{\mathcal{N}}=2$ and $\mathcal{N}=4$ supersymmetric
Yang-Mills theories. The motivation behind doing it is two-fold: (i) to establish the
anticipated integrability as well as its dependence on the number of supercharges and (ii) to
unravel the underlying integrable long-range interaction. In our previous publications
\cite{BelKorMul04} we have addressed the same questions in the sector of three-particle
gaugino operators with arbitrary number of derivatives in super Yang-Mills theories with
$\mathcal{N} = 1, 2, 4$ supercharges. We have computed the spectra of their anomalous
dimensions and have found that the energy of the states with zero quasimomentum is double
degenerate indicating the existence of higher conserved charges in addition to the quadratic
conformal Casimir \cite{Kor95,GraMat95,BraDerMan98,Bel99,BeiKriSta03}. In the present paper
we continue the analysis of noncompact sectors closed under renormalization and extend our
consideration to single-trace Wilson operators built from $L$ scalar fields $X$ and arbitrary
number of covariant derivatives,
\be
\label{Def-OpeLoc}
\mathcal{O}_{\bit{\scriptstyle n}}(0) = {\rm tr} \left\{ (i D_+)^{n_1} X (0) (i
D_+)^{n_2} X (0) \dots (i D_+)^{n_L} X (0) \right\} \, ,
\ee
where $\bit{n} = (n_1, n_2, \dots, n_L)$ denotes a set of $L$ nonnegative integers. Here
$n^\mu D_\mu = D_+ = \partial_+ - i g [A_+, \cdot]$ is the covariant derivative projected
onto the light cone with the help of the light-like vector $n^\mu$. This projection
automatically selects the maximal spin component. In order to avoid mixing with operators
built from gauginos and gluons, we choose all $X$'s to be the same and to possess the maximal
charge with respect to the internal $R$-symmetry group, that is, $X = \phi$ in $\mathcal{N}=2$
theory and $X = \phi_1 + i \phi_2$, conventionally called $\mathcal{Z}$, in $\mathcal{N} = 4$
theory. The scalar operators \re{Def-OpeLoc} carry the Lorentz spin $N= n_1 + \dots +n_L$ and
the canonical dimension $N+L$ and they can mix under renormalization with operators having
the same $N$ and $L$.

The paper is organized as follows. In Section \ref{TwoLoopFeynman} we employ a well-developed
QCD technique for perturbative computation of the dilatation operator in the momentum
representation. In this representation, the dilatation operator can be realized as an integral
operator acting on light-cone momenta of scalar fields. The mixing matrix in the basis of local
Wilson operators is simply obtained by forming Mellin moments of its integral kernel and the
explicit expressions can be found in Appendix. In Section \ref{Eigenspectrum} we discuss the
spectrum of anomalous dimensions for scalar operators \re{Def-OpeLoc} and describe integrable
structures behind the two-loop dilatation operator using the method of the Baxter
$\mathbb{Q}-$operator.  Section \ref{Discussion} contains concluding remarks.

\section{Two-loop noncompact dilatation operator}
\label{TwoLoopFeynman}

A concise representation of the entire tower of local Wilson operators is achieved
by means of non-local operators with elementary fields located at positions $z_i$
on the light-ray, $X (z_i n^\mu) \equiv X (z_i)$
\be\label{O-nonlocal}
\mathbb{O} (\bit{z}) = {\rm tr} \left\{ X (z_1) [z_1, z_2] X (z_2)[z_2,z_3] \dots
X (z_L) [z_L, z_1] \right\} \, ,
\ee
where $\bit{z} = (z_1, z_2, \dots, z_L)$ and the Wilson line  $[z_j, z_{j + 1}] =
i g \int_{z_{j + 1}}^{z_j} dz A_+ (z)$ is stretched between two fields to make
the composite operators gauge invariant. The local Wilson operators
\re{Def-OpeLoc} are deduced from $\mathbb{O} (\bit{z})$ by means of the Taylor
expansion
\be\label{O1}
\mathbb{O} (\bit{z}) = \sum_{n_1, n_2, \dots, n_L \geq 0}^\infty \frac{(- i
z_1)^{n_1}}{n_1 !} \frac{(- i z_2)^{n_2}}{n_2 !} \dots \frac{(- i z_L)^{n_L}}{n_L
!} \mathcal{O}_{\bit{\scriptstyle n}}(0) \, .
\ee
For our purposes yet it will be extremely useful to use a representation of the
same operators in the reciprocal momentum space. It is given by the Fourier
transform of $\mathbb{O} (\bit{z})$ with respect to the light-cone coordinates
\be\label{O2}
\widetilde{\mathbb{O}} (\bit{u}) =\int_{-\infty}^\infty \frac{dz_1}{2\pi}{\rm
e}^{i u_1 z_1}\ldots\int_{-\infty}^\infty \frac{dz_L}{2\pi}{\rm e}^{i u_L z_L}
\, \mathbb{O} (\bit{z}) \, ,
\ee
with $\bit{u} = (u_1, u_2, \dots, u_L)$ being the vector of the light-cone momenta.
These operators obey the renormalization group (Callan-Symanzik) equation
\be\label{measure}
\left( \mu \frac{\partial}{\partial\mu} + \beta_{\mathcal{N}} (g)
\frac{\partial}{\partial g} \right) \widetilde{\mathbb{O}} (\bit{u}) = \int [d
\bit{v}]_L \, \mathbb{V}(\bit{u} | \bit{v}) \widetilde{\mathbb{O}} (\bit{v}) \, ,
\ee
with the integration measure $[d \bit{v}]_L=dv_1\ldots dv_L \delta (\sum_k v_k -
\sum_m u_m)$ and the dilatation operator in the momentum-space representation,
$\mathbb{V}(\bit{u} | \bit{v})$, admitting perturbative expansion in 't Hooft
coupling $\lambda = g_{\rm \scriptscriptstyle YM}^2 N_c/(8 \pi^2)$
\ba
\label{PerturbExpansionKernel}
\mathbb{V}(\bit{u} | \bit{v}) = \lambda \mathbb{V}^{(0)}(\bit{u} | \bit{v}) +
\lambda^2 \mathbb{V}^{(1)}(\bit{u} | \bit{v}) + \mathcal{O}( \lambda^3 ) \, .
\ea
A detailed account on the technique used for the perturbative calculation of
the kernels $\mathbb{V}^{(0)}(\bit{u} | \bit{v})$ and $\mathbb{V}^{(1)}(\bit{u} |
\bit{v})$ can be found in our previous publication~\cite{BelKorMul04}.

In this paper, we calculate the two-loop kernel $\mathbb{V}(\bit{u} | \bit{v})$
for nonlocal light-cone scalar operators \re{O-nonlocal} in supersymmetric
Yang-Mills theories with $\mathcal{N}=2$ and $\mathcal{N}=4$ supercharges. We
shall employ the light-cone formalism of Ref.\ \cite{BelDerKorMan04} which heavily
relies on the light-cone gauge $A_+(x) = 0$ and which allows one to treat all
supersymmetric theories in a unified fashion. Under this gauge condition, the
Wilson lines in \re{O-nonlocal} reduce to unity $[z_k, z_{k + 1}] = \1$ and the
number of relevant Feynman diagrams decreases significantly. In the light-cone
formalism, the Lagrangian  of super Yang-Mills theory depends on physical
components of elementary fields only since all non-propagating degrees of freedom
can be integrated out. This allows one to switch from covariant spinor  and
vector fields to single-component fermionic and bosonic fields carrying a
definite helicity and, then, introduce dimensional regularization inside Feynman
integrals by continuing them to $D = 4 - 2 \varepsilon$ dimensions without
breaking supersymmetry of the underlying gauge theory. The resulting
regularization procedure is equivalent to the dimensional reduction scheme.

\subsection{Evolution kernel}

Let us start with the one-loop kernel $\mathbb{V}^{(0)}(\bit{u} | \bit{v})$. In
the multi-color limit, it has a simple nearest neighbor structure
\be\label{LO}
\mathbb{V}^{(0)} (\bit{u} | \bit{v}) = \sum_{k = 1}^L \bigg\{
\mathbb{V}^{(0)}_{k, k + 1} - \Gamma^{(0)} \delta(u_k-v_k) \bigg\} \prod_{{j = 1,
\atop j \neq k, k + 1}}^L \delta(u_j-v_j) \,,
\ee
with the periodicity condition $L + 1 = 1$ and the total momentum conservation
$\sum_k v_k=\sum_m u_m$ absorbed into the integration measure in \re{measure}.
Here, $\Gamma^{(0)} = \ft12 (\mathcal{N} - 4)$ is an additive constant and the
two-particle kernel is defined as
\be\label{plus}
\mathbb{V}^{(0)}_{k,k + 1} = \left[ \Theta (u_k, v_k) f_s (u_k,
v_k)\right]_+^{(u_k)} + \left[ \Theta (u_{k + 1}, v_{k + 1}) f_s (u_{k + 1}, v_{k
+ 1}) \right]_+^{(u_{k+1})} \, ,
\ee
where the generalized step-function $\Theta(u,v)$ specifies possible values of
the momentum fractions and a notation was introduced for the decoration factor
$f_s$,
\be
\Theta (u, v) = \theta (u) \theta (u - v) - \theta (- u) \theta (v - u) \, ,
\qquad f_s (u, v) = \frac{1}{v - u} \, .
\ee
The plus-distribution in \re{plus} regularizes the end-point singularities of
$f_s(u,v)$ as $u-v\to 0$ and is conventionally defined as
\be
\label{Def-Sin+}
\left[ \frac{\tau (u)}{v - u} \right]_+^{(u)} = \frac{\tau (u)}{v - u} - \delta
(u - v) \int d w \frac{\tau (w)}{v - w} \, ,
\ee
for any test function $\tau$.

\begin{figure*}[t]
\begin{center}
\mbox{
\begin{picture}(0,75)(242,0)
\psfrag{V}[cc][cc]{$\mathbb{V}$}
\put(0,-9){\insertfig{17}{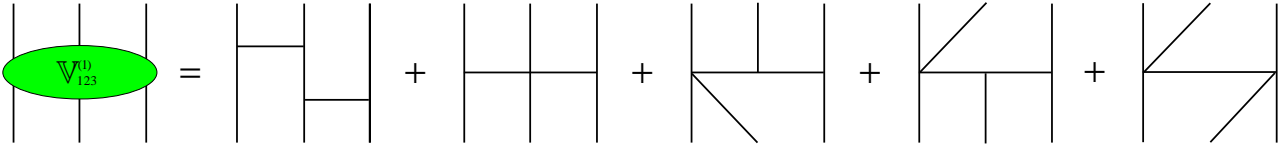}}
\end{picture}
}
\end{center}
\caption{\label{3ParticleDiag} Different topologies of two-loop diagrams in the
light-cone superspace formalism of Ref.\ \cite{BelDerKorMan04} involving three
nearest superfields. The three-particle kernel \re{Res123-ScaSecV} is obtained by
projecting the superfields onto the scalar field component. Mirror symmetric diagrams
are implied. The graphs with two-to-one and one-to-two particle transitions all
vanish.}
\end{figure*}

\begin{figure*}[t]
\begin{center}
\mbox{
\begin{picture}(0,75)(242,0)
\psfrag{V}[cc][cc]{$\mathbb{V}$}
\put(0,-9){\insertfig{17}{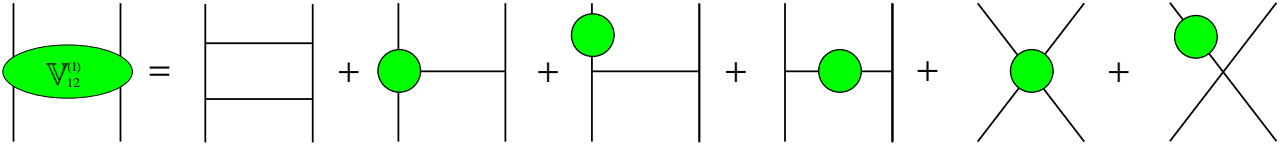}}
\end{picture}
}
\end{center}
\caption{\label{2ParticleDiag} Topologies of two-loop Feynman graphs providing
two-particle contribution to the dilatation operator. The blobs stand for
one-loop correction to the corresponding propagators and three- and four-particle
interaction vertices. Mirror symmetric diagrams have to be added.}
\end{figure*}

We now turn to the two-loop contribution to \re{PerturbExpansionKernel}. In the
multi-color limit, the interaction can involve up to three nearest neighbor fields
and the generic structure of the kernel $\mathbb{V}^{(1)} (\bit{u} | \bit{v})$
reads
\be\label{V1}
\mathbb{V}^{(1)} (\bit{u} | \bit{v}) = \sum_{k = 1}^L \bigg\{
\mathbb{V}^{(1)}_{k, k + 1, k + 2} +  \delta(u_{k}-v_{k})\left[
\mathbb{V}^{(1)}_{k+1, k + 2} - \Gamma^{(1)} \delta(u_{k+1}-v_{k+1})\right]
 \bigg\} \prod_{{j = 1, j \neq k, \atop k + 1, k + 2}}^L
\delta(u_j-v_j)\, .
\ee
The three- and two-particle kernels $\mathbb{V}_{k, k + 1, k + 2}$,
$\mathbb{V}_{k, k + 1}$ are found from the diagrams displayed in
Figs.\ \ref{3ParticleDiag} and \ref{2ParticleDiag}, respectively. Finally,
$\Gamma^{(1)}$ receives contributions from two-loop renormalization of the
external scalar legs. We will not dwell on the computation procedure since it was
spelled out in details in Ref.\ \cite{BelKorMul04}.

The calculation of the Feynman graphs in Fig.\ \ref{3ParticleDiag} yields the
three-particle kernel in $\mathcal{N}=2$ and $\mathcal{N}=4$ SYM theories as
\ba
\label{Res123-ScaSecV}
\mathbb{V}^{(1)}_{1 2 3} = \!\!\!&-&\!\!\! \frac12 \left[ \Theta^{(1)}_{123} f_s
(u_1, v_1) f_s (u_{3}, v_{3}) \ln\frac{u_1 (u_1 - v_1)^2}{v_1 (u_{3} - v_{3} -
v_{2}) u_{2}} \right]_+^{(u_{3})}
\nonumber\\
&+&\!\!\! \frac{1}{2} \left[ \Theta^{(4)}_{123} f_s (u_1, v_1) f_s (u_{3}, v_{3})
\ln\frac{(v_{1}+v_{3} - u_1)(v_{2}+v_{3} - u_{3}) }{u_{2} v_{2}}
\right]_{++}^{(u_1 u_3)}
\nonumber\\
&+&\!\!\! +\left\{u_1 \leftrightarrow u_{3} \atop  v_1 \leftrightarrow v_{3}
\right\} \, .
\ea
In general, the other kernels $\mathbb{V}^{(1)}_{k,k+1,k+2}$ can be obtained from
this expression through cyclic permutation of indices under the periodic boundary
condition $L+k\equiv k$. Here the generalized three-particle step-functions are
defined following the conventions of Ref.\ \cite{BelKorMul04} as
\ba
\Theta^{(1)}_{123} \!\!\!&=&\!\!\! \Theta (u_1, v_1) \left[ \Theta (u_{2},
v_{1}+v_{2} - u_1 ) - \Theta (v_{2}+v_{3} - u_{3}, v_{3}) \right]
\, , \\
\Theta^{(4)}_{123} \!\!\!&=&\!\!\! \Theta (u_{2}, v_{2}+v_{3} - u_{3}) \Theta
(v_{2}+v_{3} - u_{3}, v_{2}) \, .
\ea
For $v_{1,2,3} \ge 0$ and $v_{1}+v_{2}+v_{3}=1$ they define the regions in
$(u_1,u_3)-$plane (with $u_2=1-u_1-u_3$) which produce non-vanishing
contributions to $\mathbb{V}^{(1)}_{1 2 3}$:
\be
\Theta^{(1)}_{123}
=
\left[ {0 \leq u_1  \leq v_1 \atop v_{2}+v_{3} \leq  u_{3} \leq 1-u_{1}} \right]
, \qqqquad
\Theta^{(4)}_{123}
=
\left[ {v_1 \leq u_1  \leq 1-u_{3} \atop v_{3} \leq u_{3} \leq 1} \right]
.
\ee
Remarkably, the scalar kernel \re{Res123-ScaSecV} can be obtained from the
three-particle gaugino kernel (see Eqs.~(5.7)-(5.14) in \cite{BelKorMul04})
by substituting the gaugino decorating factor $f_q (u,v) = u/(v(v-u))$ with
the scalar one $f_s (u,v)$.

In Eq.\ (\ref{Res123-ScaSecV}), in the first term the superscript $(u_3)$ indicates
that the plus-distribution is taken only with respect to the variable $u_{3}$ as
shown above in Eq.\ (\ref{Def-Sin+}). The second term in \re{Res123-ScaSecV} involves
the double plus-prescription defined as
\baa
\left[ \frac{\tau (u, u')}{(v - u)(v' - u')} \right]_{++}^{(u u')}
\!\!\!&=&\!\!\! \frac{\tau (u, u')}{(v - u)(v' - u')} + \delta (u - v) \delta (u'
- v') \int d w \, d w' \frac{\tau (u, w')}{(v - u)(v' - w')}
\\
&-&\!\!\!
\delta (u - v)
\int d w \frac{\tau (w, u')}{(v - w)(v' - u')}
-
\delta (u' - v')
\int d w' \frac{\tau (w, w')}{(v - w)(v' - w')}
\, , \nonumber
\eaa
with $\tau (u, u')$ being a test function.

The two-particle kernel $\mathbb{V}^{(1)}_{k,k+1}$ receives contributions from
the diagrams presented in Fig.\ \ref{2ParticleDiag} as well as from subtraction
terms coming from the single and double plus-distributions. The expression for
$\mathbb{V}^{(1)}_{12}$ looks like
\be
\label{Res-12-ScaSec-NLO}
\mathbb{V}^{(1)}_{12} =  \left[\Theta (u_1, v_1) f_s (u_1, v_1) \left( 4 -
\mathcal{N} - \frac{\pi^2}{6} + \frac{1}{4}\ln^2\frac{u_1}{v_1} -
\ln\frac{u_1}{v_1} \ln\left(1 - \frac{u_1}{v_1} \right) \right)\right]_+^{(u_1)}
+ \left\{ u_1 \leftrightarrow u_{2} \atop  v_1 \leftrightarrow v_{2} \right\} \,
.
\ee
As before, all other $\mathbb{V}^{(1)}_{k,k+1}$ can be obtained from
$\mathbb{V}^{(1)}_{12}$ by cyclically permuting the indices. Finally, one has to
add a single-particle contribution due to two-loop renormalization of the scalar
fields. The former can be read off from the renormalization constant of the
gauginos in the dimensional reduction scheme discussed at length in Ref.\
\cite{BelKorMul04}
\be
\label{Gamma1}
\Gamma^{(1)} = \frac{1}{2} (\mathcal{N} - 4) (2 - \mathcal{N}) \, ,
\ee
so that $\Gamma^{(1)}=0$ in $\mathcal{N}=2$ and $\mathcal{N}=4$ theories. This
result has a simple interpretation (see Eq.~\re{local} below).

It is interesting to notice that the three-particle kernel $\mathbb{V}^{(1)}_{123}$
does not depend on the number of supercharges $\mathcal{N}$. At the same time,
examining the $\mathcal{N}-$dependence of the two-loop kernel $\mathbb{V}^{(1)}_{12}$
one finds the following remarkable relation between the two-loop dilatation operators
in $\mathcal{N} = 2$ and $\mathcal{N} = 4$ theories,
\be\label{N2-4}
\mathbb{V}_{\mathcal{N} = 2}(\bit{u} | \bit{v}) = \left( 1 + 2 \lambda \right)
\mathbb{V}_{\mathcal{N} = 4}(\bit{u} | \bit{v})\,.
\ee
The same relation was previously found for the gaugino operators in Ref.~\cite{BelKorMul04}.
It suggests that the dilatation operators in the two theories share the same properties
to two loops at least, and their eigenvalues are, obviously, related via Eq.\ \re{N2-4}.

Being combined together, Eqs.~\re{PerturbExpansionKernel}, \re{V1},
\re{Res123-ScaSecV}, \re{Res-12-ScaSec-NLO} and \re{Gamma1} provide explicit
expressions for the two-loop dilatation operator for the scalar operators
\re{O-nonlocal} in $\mathcal{N}=2$ and $\mathcal{N}=4$ SYM theories in the
multi-color limit. In section \ref{Eigenspectrum}, we will diagonalize the
dilatation operator \re{PerturbExpansionKernel} and determine the spectrum of
anomalous dimensions of scalar operators. To accomplish this goal however, we
will first reconstruct the mixing matrix for the operators \re{Def-OpeLoc}.

\subsection{Mixing matrix}
\label{SubSec-MixMat}

It follows from \re{O1} and \re{O2} that the local Wilson operators \re{Def-OpeLoc}
are related to the moments of the operators in the momentum representation
$\widetilde{\mathbb{O}} (\bit{v})$
\be
\mathcal{O}_{\bit{\scriptstyle n}}(0) = \int dv_1 v_1^{n_1} \ldots \int dv_L
v_L^{n_L}\,\widetilde{\mathbb{O}} (\bit{v})\,.
\ee
Together with \re{measure} this implies that the operators
$\mathcal{O}_{\bit{\scriptstyle n}}(0)$ obey the renormalization group equation
\be\label{measure1}
\left(
\mu \frac{\partial}{\partial\mu} + \beta_{\mathcal{N}} (g) \frac{\partial}{\partial g}
\right) \mathcal{O}_{\bit{\scriptstyle n}}(0)
= - \sum_{\bit{\scriptstyle m}}
{\Lambda}^{\bit{\scriptstyle m}}_{\bit{\scriptstyle n}}(\lambda)\,
\mathcal{O}_{\bit{\scriptstyle m}}(0)
\ee
with $ \bit{m}=(m_1,\ldots,m_L)$ such that $m_i\ge 0$ and $\sum_k n_k = \sum_k
m_k$. The mixing matrix is introduced as
\be
\label{Def-MelMom-V} \int [d \bit{u}]_L \, u^{n_1}_1 u^{n_2}_2 \dots u^{n_L}_L \,
\mathbb{V} (\bit{u} | \bit{v}) = - \sum_{\bit{\scriptstyle m}}
\Lambda^{\bit{\scriptstyle m}}_{\bit{\scriptstyle n}} (\lambda)
v^{m_1}_1 v^{m_2}_2 \dots v^{m_L}_L \, .
\ee
It defines the representation of the evolution kernel $\mathbb{V} (\bit{u} |
\bit{v})$ in the space spanned by the polynomials $v^{n_1}_1 v^{n_2}_2 \dots
v^{n_L}_L$. Substituting the evolution kernel in \re{Def-MelMom-V} by its
two-loop expression \re{PerturbExpansionKernel}, we obtain the corresponding
mixing matrix in the multi-color limit. In particular, for $n_1=\ldots=n_L=0$
one finds that the terms in \re{PerturbExpansionKernel} containing `+' and `++'
distributions provide vanishing contribution to the left-hand side of
\re{Def-MelMom-V} leading to
\be\label{local}
\Lambda^{\bit{\scriptstyle 0}}_{\bit{\scriptstyle 0}} (\lambda)= L
\left[\lambda\Gamma^{(0)} + \lambda^2\Gamma^{(1)} + \mathcal{O}(\lambda^3)\right]
=\frac{L}2 \left[ \lambda(\mathcal{N} - 4) + \lambda^2(\mathcal{N} - 4) (2 -
\mathcal{N})+ \mathcal{O}(\lambda^3) \right]
\ee
According to \re{measure1} and \re{Def-MelMom-V}, $\Lambda^{\bit{\scriptstyle 0}}_{
\bit{\scriptstyle 0}} (\lambda)$ defines the anomalous dimension of the local operator
$\mathbb{O} (0) = {\rm tr} \, [X^L (0)]$ in $\mathcal{N} = 2$ and $\mathcal{N}=4$
theories. These operators are known to be protected~\cite{MagTan01} and, therefore,
their anomalous dimension is equal, in our notations, to $L\beta_{\mathcal{N}} (g)/2$.
This implies that \re{local} is exact to all loops.

The perturbative expansion of the mixing matrix $\Lambda^{\bit{\scriptstyle m}}_{
\bit{\scriptstyle n}}(\lambda)$ is similar to that of the evolution kernel
\re{PerturbExpansionKernel} and reads in the multi-color limit
\be\label{mix-mat}
\Lambda^{\bit{\scriptstyle m}}_{ \bit{\scriptstyle n}}(\lambda)
=
\lambda \Lambda^{(0) \bit{\scriptstyle m}}_{\phantom{(0)} \bit{\scriptstyle n}}
+
\lambda^2 \,
\Lambda^{(1) \bit{\scriptstyle m}}_{\phantom{(1)} \bit{\scriptstyle n}}
+
\mathcal{O}(\lambda^2)
\ee
with $\lambda=g^2N_c/(8\pi^2)$. The matrices ${\Lambda^{(0)}}$ and ${\Lambda^{(1)}}$ are
given by the moments \re{Def-MelMom-V} of the kernels $\mathbb{V}^{(0)}(\bit{u} | \bit{v})$,
Eq.~\re{LO}, and $\mathbb{V}^{(1)}(\bit{u} | \bit{v})$, Eq.~\re{V1}, respectively,
\ba
\Lambda^{(0) \bit{\scriptstyle m}}_{\phantom{(0)} \bit{\scriptstyle n}}
\!\!\!&=&\!\!\
\sum_{k = 1}^L
\left\{
\Lambda^{(0) m_k m_{k + 1}}_{\phantom{(0)} n_k n_{k + 1}}
+
\Gamma^{(0)}
\delta^{m_k}_{n_k} \delta^{m_{k + 1}}_{n_{k + 1}} \right\}
\prod_{j = 1, \atop j \neq k, k + 1}^L \delta^{m_j}_{n_j}
\, ,
\\ \label{Lambda}
\Lambda^{(1) \bit{\scriptstyle m}}_{\phantom{(1)} \bit{\scriptstyle n}}
\!\!\!&=&\!\!\!
\sum_{k = 1}^L
\left\{ \Lambda^{(1) m_k m_{k + 1} m_{k + 2}}_{\phantom{(1)} n_k n_{k + 1} n_{k +
2}} + \Lambda^{(1) m_k m_{k + 1}}_{\phantom{(1)} n_k n_{k + 1}} \delta^{m_{k +
2}}_{n_{k + 2}} + \Gamma^{(1)} \delta^{m_k}_{n_k} \delta^{m_{k + 1}}_{n_{k + 1}}
\delta^{m_{k + 2}}_{n_{k + 2}} \right\} \prod_{j = 1, j \neq k,\atop  k + 1, k +
2}^L \delta^{m_j}_{n_j} \, . \nonumber
\ea
Here $\delta_{n}^{m}$ is the Kronecker symbol and periodic boundary conditions
$L+k=k$ are implied. We also introduced a notation for the moments of the two-
and three-particle kernels. The explicit expressions for the corresponding
matrices $\Lambda^{(0) m_1 m_2}_{\phantom{(0)} n_1 n_2}$, $\Lambda^{(1) m_1
m_2}_{\phantom{(1)} n_1 n_2}$ and $\Lambda^{(1) m_1 m_2 m_3}_{\phantom{(1)} n_1
n_2 n_3}$ are rather lengthy and can be found in Appendix.

\section{Eigenspectrum and integrability}
\label{Eigenspectrum}

To solve the evolution equation \re{measure1}, one examines the eigenproblem for
the mixing matrix \re{mix-mat}
\be\label{aux}
\sum_{\bit{\scriptstyle n}}
\Lambda^{\bit{\scriptstyle m}}_{ \bit{\scriptstyle n}} (\lambda)\, \Psi(\bit{n})
=
\gamma(\lambda)\, \Psi(\bit{m})
\, ,
\ee
where the sum runs over $\bit{n}=(n_1,\ldots,n_L)$ such that $n_i \ge 0$ and $\sum_k
n_k = \sum_k m_k$. We remind that the $\beta-$function vanishes in the $\mathcal{N}=4$
SYM theory $\beta_{\mathcal{N}=4}(g)=0$, while its exact value in the $\mathcal{N}=2$
SYM theory is given by the one-loop expression $\beta_{\mathcal{N}=2}(g)=-2\lambda$.
Then, it follows from Eq.\ \re{measure1} that in ${\mathcal{N}=4}$ theory, the conformal
operators $\mathcal{O}_{\rm conf}(0)=\sum_{\bit{\scriptstyle n}} \Psi({\bit{n}})
\mathcal{O}_{\bit{\scriptstyle n}}(0)$ have an autonomous scale dependence with
$\gamma(\lambda)$ being the corresponding anomalous dimensions. In $\mathcal{N}=2$
theory the scale dependence of the operators $\mathcal{O}_{\rm conf}(0)$ is more
involved due to the dependence of the eigenstates $\Psi({\bit{n}})$ on the running
coupling constant but one still refers to $\gamma(\lambda)$ as an anomalous dimension.

Having the explicit expression for the two-loop mixing matrix \re{mix-mat} at our
disposal, we can solve the spectral problem \re{aux} for arbitrary lengths $L$
and look for manifestation of symmetries of the dilatation operator in its
spectrum. The evolution kernel $\mathbb{V} (\bit{u} | \bit{v})$ preserves the
total momentum $P=\sum_n u_n=\sum_k v_k$ and its eigenvalues $\gamma(\lambda)$ do
not depend on $P$. This allows one to simplify the analysis by going over to the
forward limit, $P=0$, see Ref.\ \cite{BelKorMul04}. Still, the anomalous
dimensions $\gamma(\lambda)$ depend on the total number of derivatives
$N=\sum_{i=1}^L n_i$ which is one of the integrals of motion for the
Schr\"odinger-like equation \re{aux}. Another conserved charge follows from the
invariance of the mixing matrix \re{mix-mat} under discrete cyclic $\mathbb{P}$
and mirror transformations $\mathbb{M}$ defined as
\ba
\label{DiscreteSymmetries} \mathbb{P}\, \Psi (n_1, n_2, \dots, n_L) &=& \Psi
(n_2, n_3, \dots, n_1) \, , \qquad
\\ \nonumber
\mathbb{M} \, \Psi (n_1, n_2, \dots, n_L) &=& \Psi (n_L, n_{L - 1}, \dots, n_1)
\, .
\ea
Since these two operators do not commute with each other, the solutions to \re{aux}
can be classified according to the eigenvalues of only one of them, say
$\mathbb{P}$
\be\label{quasi}
\mathbb{P} \Psi (n_1, n_2, \dots, n_L)
=
{\rm e}^{i \theta} \Psi (n_1, n_2, \dots, n_L)
\, ,
\ee
where the quasimomentum $\theta$ takes $L$ distinct values, $\theta = \frac{2 \pi n}{L}$
with $n = 0, 1, \dots, L - 1$. Making use of the relation $\mathbb{P} \mathbb{M} \mathbb{P}
= \mathbb{M}$, one immediately finds from \re{aux} that for $\theta \neq 0$ the eigenstates
$\Psi(\bit{n})$ and $\mathbb{M}\,\Psi(\bit{n})$ have the same ``energy'' $\gamma(\lambda)$
and opposite values of the quasimomentum, $\theta$ and $-\theta$, respectively. Thus, the
solutions to \re{aux} with nonzero quasimomentum are necessarily double degenerate. For
the eigenstates with $\theta=0$ the discrete symmetry alone does not imply any degeneracy.
We recall that the eigenstates $\Psi(\bit{n})$ determine the form of conformal operators
$\mathcal{O}_{\rm conf}(0)=\sum_{\bit{\scriptstyle n}} \Psi(\bit{n})
\mathcal{O}_{\bit{\scriptstyle n}} (0)$ with the basis operators
$\mathcal{O}_{\bit{\scriptstyle n}}(0)$ defined in \re{Def-OpeLoc}. Since the latter
operators are cyclically symmetric with respect to $\bit{n}$, the eigenstates $\Psi(\bit{n})$
should possess the same symmetry, that is, they ought to have zero quasimomentum $\theta=0$.
This leads to an additional selection rule for solutions to Eq.\ \re{aux}. We would like to
stress that the mixing matrix \re{mix-mat} possesses eigenvalues with both quasimomenta
$\theta\neq 0$ and $\theta=0$ but only the latter define eigenvalues of the dilatation
operator in gauge theory.

\begin{figure}[t]
\begin{center}
\mbox{
\begin{picture}(480,135)
\put(-3,-15){\insertfig{5.5}{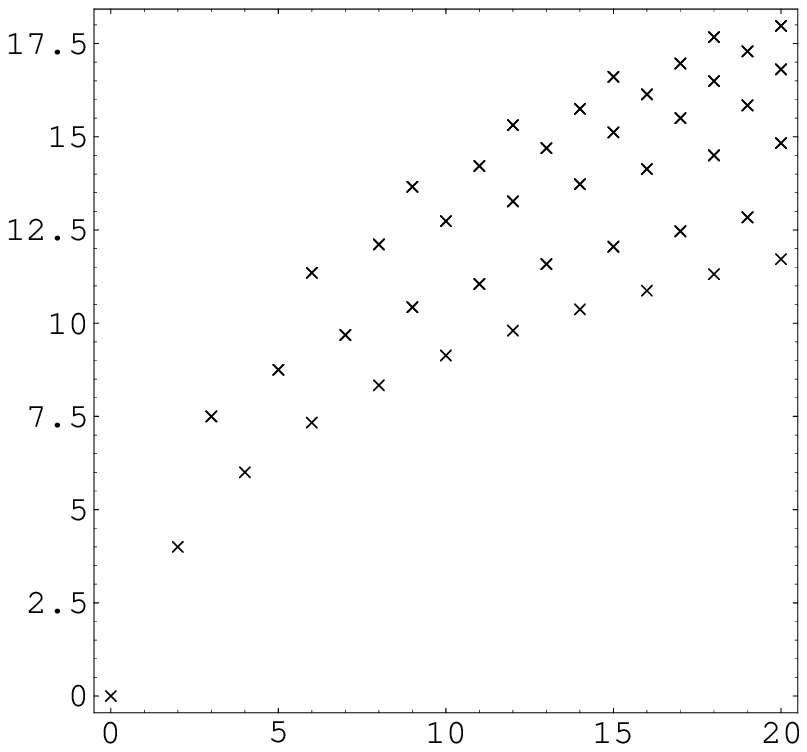}}
\put(25,120){(a)}
\put(170,-10){\insertfig{5.17}{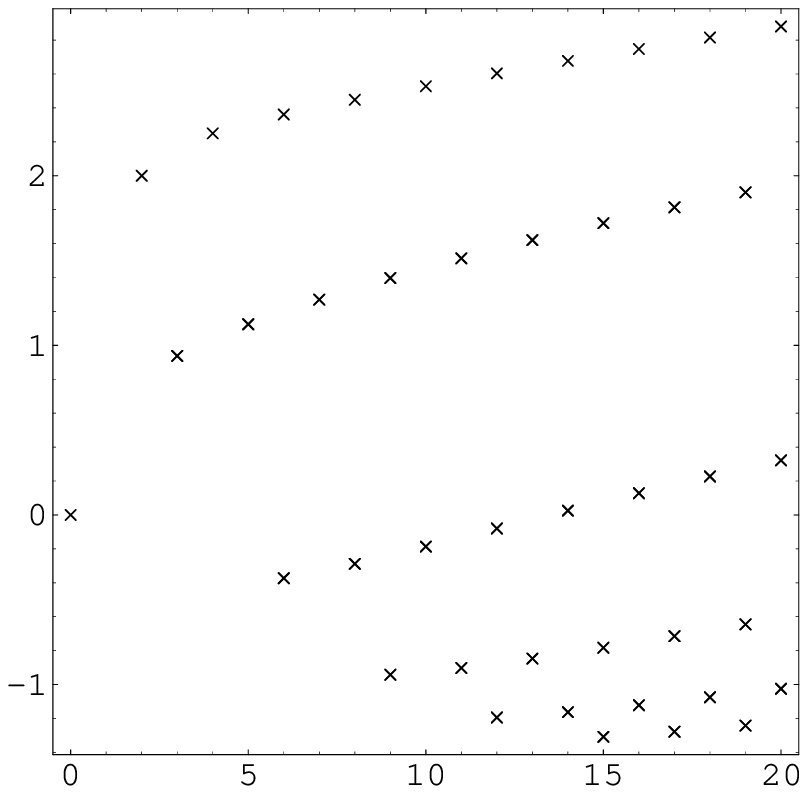}}
\put(190,120){(b)}
\put(335,-12){\insertfig{5.3}{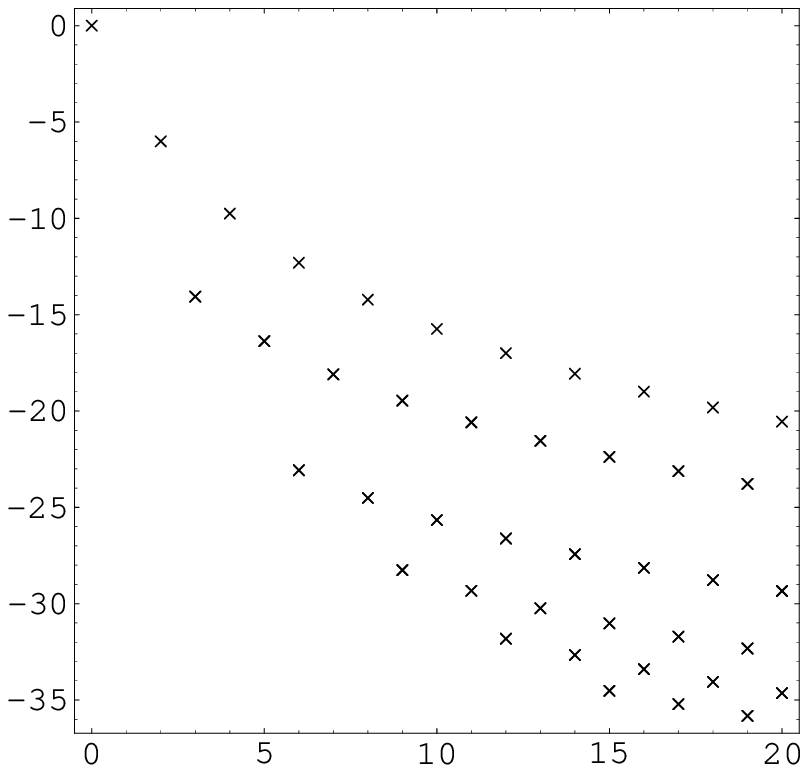}}
\put(465,120){(c)}
\end{picture}
}
\end{center}
\caption{\label{SpectrumAD} The spectrum of anomalous dimensions of $L = 3$ operators
(see text) at one loop ($a$), and two loops for $\mathcal{N} = 2$ ($b$) and $\mathcal{N}
= 4$ ($c$) super Yang-Mills theories.}
\end{figure}

To two-loop accuracy, the anomalous dimensions in the multi-color limit have the perturbative
expansion \be\label{gamma} \gamma (\lambda) = \lambda \, \varepsilon^{(0)} + \lambda^2
\,\varepsilon^{(1)} + \mathcal{O} (\lambda^3) \, , \ee with $\varepsilon^{(0)}$ and
$\varepsilon^{(1)}$ being functions of the length of the operator $L$, the total number of
derivatives $N$ and some other quantum numbers, yet to be determined. To find the explicit
form of these functions one has to diagonalize the mixing matrix for various $L$ and $N$.
In particular, using the expression for the two-loop mixing matrix, Eqs.~\re{mix-mat} and
\re{Lambda}, and solving the eigenproblem \re{aux} for scalar operators of length $L=3$ and
the total number of derivatives $0\le N \le 20$, we calculated the values of $\varepsilon^{(0)}$
and $\varepsilon^{(1)}$ with $\theta=0$ in the $\mathcal{N}=2$ and $\mathcal{N}=4$ SYM theories.
We summarized our results in Fig.\ \ref{SpectrumAD}. We found that in both theories all eigenvalues
\re{gamma} (except of a single lowest eigenvalue for each even $N$) are double degenerate to two
loops . We would like to stress that the two-loop evolution kernel in the $\mathcal{N}=2$ theory
contains conformal symmetry breaking terms proportional to the $\beta-$function. In the same
fashion as in Ref.\ \cite{BelKorMul04}, the fact that the degeneracy is present in the
$\mathcal{N}=2$ theory suggests that the phenomenon is not directly tied to the conformal
symmetry.

\subsection{One loop}

We recall that for eigenstates with vanishing quasimomentum $\theta=0$, the discrete
symmetry \re{DiscreteSymmetries} is not sufficient to warrant the double degeneracy of the
spectrum. To one-loop order, this property immediately follows from hidden integrability
of the one-loop dilatation operator~\cite{Kor95,BraDerMan98,Bel99,BeiKriSta03}. Namely,
the one-loop mixing matrix $\Lambda^{(0)}$ can be mapped into a Hamiltonian of the XXX
Heisenberg spin chain of length $L$ and spin operators in all sites being generators of
(infinite-dimensional) discrete series representation of the $SL(2; \mathbb{R})$ group
of spin $j=\ft12$. On the gauge theory side, this group emerges as the so-called collinear
subgroup of the full (super)conformal group. Its representation is determined by the
conformal spin of scalar fields entering the Wilson operators \re{Def-OpeLoc}. It is
important that for the one-loop dilatation operator, the conformal spin of scalar field
can be substituted by its value at order $\lambda^0$, that is, by its classical value
$j=\ft12$. As we will argue below the situation becomes more complex starting from two
loops.

Thanks to integrability, the spectral problem \re{aux} can be solved exactly to
one-loop order using the Quantum Inverse Scattering Method \cite{TakFad79}. This
method allows us to identify the complete set of conserved charges $q_2^{(0)},
\ldots, q_L^{(0)}$ whose eigenvalues parameterize solutions to \re{aux}. For scalar
operators of length $L$, they can be obtained as coefficients in the expansion of the
auxiliary transfer matrix in powers of the spectral parameter
$u$
\be
\label{t} t_L^{(0)} (u) = {\rm tr} \left\{ \mathbb{L}_L (u) \mathbb{L}_{L - 1}
(u) \dots \mathbb{L}_1 (u) \right\}
= u^L \left[2 + q_2^{(0)}u^{-2} + q_3^{(0)}u^{-3} + \dots + q_L^{(0)}u^{-L}
\right]\, .
\ee
Here $\mathbb{L}_k (u)=\II\cdot u + i \bit{\sigma} \cdot \bit{S}_k$ is the
standard Lax operator for the XXX Heisenberg spin chain and ${\sigma^a}$ are
Pauli matrices. It is given by a $2\times 2$ matrix whose entries are spin
operators $S_k^a$ in $k^{\rm th}$ site. In gauge theory, the latter are
generators of the collinear $SL(2)$ subgroup acting on $k^{\rm th}$ scalar field
in \re{Def-OpeLoc}. The ``lowest'' integral of motion $q_2^{\scriptscriptstyle
(0)}$ is related to the total conformal spin of the Wilson operator, $J=N + L/2$,
\be\label{q20}
q_2^{(0)} = - (N + \ft 12L) (N + \ft 12L  - 1) - \ft14 L  \, .
\ee
The explicit form of higher conserved charges $q_{k>2}^{(0)}$ can be found in
\cite{Kor95}. In particular, they possess a definite parity with respect to the
discrete transformations \re{DiscreteSymmetries}, $[\mathbb{P}, q^{(0)}_n] = \{
\mathbb{M}, q^{(0)}_{2k + 1} \} = [ \mathbb{M}, q^{(0)}_{2k} ] = 0$. This implies
that the one-loop ``energy'' $\varepsilon^{(0)}$ as a function of the conserved
charges $q_{k}^{(0)}$ satisfies the relation
\be\label{par}
\varepsilon^{(0)} \big( q_3^{(0)}, q_4^{(0)}, \dots, q_L^{(0)} \big) =
\varepsilon^{(0)} \big( - q_3^{(0)}, q_4^{(0)}, \dots, (- 1)^L q_L^{(0)} \big)
\, ,
\ee
and, therefore, all eigenstates including those with zero quasimomentum states
are double degenerate provided that $q^{(0)}_{2 k + 1} \neq 0$. By explicit
diagonalization of the dilatation operator in $\mathcal{N} = 2$ and
$\mathcal{N}=4$ SYM, evaluated in the previous section, we found that its
eigenspectrum \re{gamma} is double degenerate not only at one loop but also in
two-loop order. Together with \re{par} this suggests that the all-loop anomalous
dimension \re{gamma} is a function of the conserved charges $q_{k}(\lambda)$
\be\label{q-charges}
q_k (\lambda) = q_{k}^{(0)} + \lambda \,q_{k}^{(1)}+ \mathcal{O}(\lambda^2)
\, ,
\ee
such that it verifies the same parity relation \re{par}. We shall argue in
Sect. \ref{ThreeLoopBaxter} that this is indeed the case to two loops, at least.

The eigenspectrum of the one-loop anomalous dimension $\varepsilon^{(0)}$ can be
easily found using the method of the Baxter $\mathbb{Q}-$operator for the
$SL(2;\mathbb{R})$ Heisenberg spin chains. The method relies on the existence of
a commuting family of operators $\mathbb{Q}(u)$ which depend on an arbitrary
spectral parameter $u$ and, in addition, commute with the Hamiltonian (i.e., the
one-loop dilatation operator). The one-loop anomalous dimensions $\varepsilon^{(0)}$
and the corresponding quasimomenta $\theta$, Eq.~\re{quasi}, can be expressed in
terms of the eigenvalues of Baxter operator that we shall denote as $Q^{(0)}(u)$
\be
\label{OneLoopEnergy}
\varepsilon^{(0)}
=
i \left( \ln Q^{(0)} (\ft{i}{2}) \right)' - i \left( \ln Q^{(0)} (- \ft{i}{2}) \right)'
\, , \qquad
{\rm e}^{i \theta} = \frac{Q^{(0)} (\ft{i}{2})}{Q^{(0)} (- \ft{i}{2})}
\, ,
\ee
For the $SL(2;\mathbb{R})$ magnet of spin $j$ and length $L$ the function
$Q^{(0)}(u)$ satisfies the second order finite-difference equation \cite{Bax72},
the so-called ``$tQ$'' or Baxter equation
\be\label{Bax0}
\Delta^{(0)}_+(u)  Q^{(0)} (u + i) + \Delta^{(0)}_-(u) Q^{(0)} (u - i) =
t^{(0)}_L (u)\, Q^{(0)} (u) \, ,
\ee
where the ``dressing factors'' $\Delta^{(0)}_\pm(u) = (u\pm ij)^L$ depend on the
spin $j$ and $t^{(0)}_L (u)$ is a polynomial in $u$ of degree $L$ defined in
\re{t}. We recall that for the scalar operators \re{Def-OpeLoc}, to one-loop
order, the spin $j$ is given by the conformal spin of scalar field at $\lambda^0$
order, i.e., $j = \ft12$. The Baxter equation \re{Bax0} alone does not specify
$Q^{(0)} (u)$ uniquely and it has to be supplemented by an additional condition
that $Q^{(0)} (u)$ should be a polynomial in $u$ of degree $N \ge 0$~\cite{Kor95}.
Then, $Q^{(0)} (u)$ can be parameterized (modulo an overall normalization) by its roots
\be\label{Q0}
Q^{(0)}(u) = \prod_{k=1}^N (u-u_k^{(0)}) \, .
\ee
It is known that for the $SL(2;\mathbb{R})$ spin chain the roots $u_k^{(0)}$ take
real values only~\cite{Kor95}. In gauge theory, the nonnegative integer $N$ coincides
with the total number of derivatives in \re{Def-OpeLoc}. Being combined together,
Eqs.\ \re{Bax0} and \re{Q0} uniquely define $Q^{(0)}(u)$ and allow one to calculate
the one-loop anomalous dimensions \re{OneLoopEnergy} (see Fig.~\ref{SpectrumAD}a)
and determine the corresponding values of the conserved charges $q_k^{(0)}$
entering \re{t}.

The method of the Baxter $\mathbb{Q}-$operator is equivalent to the Bethe Ansatz.
Indeed, substituting $u=u_k^{(0)}$ into both sides of \re{Bax0}, one finds that
$u_k^{(0)}$ satisfy the Bethe-root equations. In the same manner, substitution of
\re{Q0} into \re{OneLoopEnergy} leads to the well-known expressions for energy it
terms of Bethe roots. In the next section, we will discuss a ``deformation'' of
the one-loop Baxter equation \re{Bax0} in order to accommodate the two-loop
corrections to the anomalous dimensions that were computed in
Sect.~\ref{TwoLoopFeynman}. The reason why we prefer to deal with the Baxter
$\mathbb{Q}-$operator is that its eigenvalues have a direct physical meaning
which should be preserved to all loops -- for real $u$ the function $Q(u)$
determines the wave function of the spin chain in separated (``collective'')
variables \cite{Skl90,DerKorMan02} (see also discussion in Sect.~4). Defined in
this way, $Q(u)$ should oscillate on the real $u-$axis and the number of its
nodes should be equal to the excitation number $N$.

\subsection{Two loops and beyond}
\label{ThreeLoopBaxter}

The double degeneracy of the spectrum of two-loop anomalous dimension established
in Sect.~\ref{TwoLoopFeynman} combined with the exact integrability of the one-loop
spectrum suggests to generalize the Baxter equation to higher loops. Going over
to two loops, we expect that the Bethe roots will be corrected by perturbative
corrections
\be\label{u-par}
u_k(\lambda) = u_k^{(0)} + \lambda\, u_k^{(1)} + \mathcal{O}(\lambda^2)
\ee
and $u_k$ will verify ``modified'' Bethe equations. For the scalar operators
under consideration, such equations have been conjectured in \cite{Sta04,BeiStau05}
\be\label{Bethe-x}
\lr{\frac{x_k^+}{x_k^-}}^L
=
\prod_{j\neq k}^N \frac{x_k^- - x_j^+}{x_k^+ - x_j^-}
\frac{1 - \lambda/(2x_k^+ x_j^-)}{1-\lambda/(2x_k^- x_j^+)}
\, ,
\ee
where $x_k^\pm = x(u_k\pm \ft{i}2)$ and the deformed spectral parameter $x=x(u)$ is
defined as \cite{BeiDipStau04}
\be
\label{x(u)}
x(u) =\ft12 u \left[ 1+ \sqrt{1-2\lambda/u^2}\right]
\, .
\ee
Then, the anomalous dimension and the corresponding quasimomentum are determined
in terms of $x_k^\pm$ parameters as follows \cite{Sta04,BeiStau05}
\be\label{Bethe1}
\gamma(\lambda) = \lambda\sum_{k=1}^N
\left(\frac{i}{x_k^+} - \frac{i}{x_k^-}\right)
\, , \qquad
\e^{i\theta} = \prod_{k=1}^N \frac{x_k^+}{x_k^-}
\, .
\ee
This relation coincides with \re{OneLoopEnergy} to one loop and it is believed
that it should reproduce the anomalous dimensions of scalar operators
\re{Def-OpeLoc} in the $\mathcal{N}=4$ SYM theory to three loops at least.

As was already mentioned above, the one-loop Bethe equations for the parameters
$u_k^{(0)}$ are equivalent to the Baxter equation \re{Bax0} for the polynomial
$Q(u)$, Eq.~\re{Q0}. We assume that the same relation between the Bethe Ansatz
and the Baxter equation also holds in higher loops and introduce into
consideration a polynomial with roots given by parameters $u_k$, Eq.~\re{u-par}
\be\label{Q-dressed}
Q (u) = \prod_{k = 1}^N (u - u_k (\lambda))
= Q^{(0)} (u)
+ \lambda\, Q^{(1)} (u)
+ \lambda^2 Q^{(2)} (u) + \dots \, .
\ee
Here the leading term $Q^{(0)} (u)$ coincides with the solution to the Baxter
equation \re{Bax0} and is given by a polynomial of degree $N$ with real roots
$u_k^{(0)}$. By construction, the subleading terms $Q^{(n)}(u)$ do not depend on
the coupling constant and are given by polynomials of degree $N - 1$, that is,
$Q^{(n)} (u) \sim u^{L - 1}$ for $n \geq 1$.

It turns out that the function $Q(u)$ defined in \re{Q-dressed} obeys a second-order
finite difference equation very similar to the Baxter equation
\re{Bax0}
\be \label{ThreeLoopBaxterEq}
\Delta_+\big( x(u+\ft{i}2) \big) Q (u + i) + \Delta_-\big( x(u-\ft{i}2) \big) Q
(u - i) = t_L \left(x(u)\right)\, Q (u) \, ,
\ee
where the dressing factors $\Delta_\sigma(x)$ (with $\sigma=\pm$) satisfy the
condition $\Delta_-(x) = \widebar{\Delta_+(x^*)}$ and read to three-loop accuracy
\be\label{Delta-dressed}
\Delta_\sigma(x) = x^L \Delta_\sigma^{\rm (ren)}  ( x ) \, ,
\ee
with the function $x=x(u)$ defined in \re{x(u)} and
\be
\label{Delta-ren} \Delta_\sigma^{\rm (ren)}(x) = \exp\left( - \frac{\lambda}{x}
\left( \ln Q (\ft{i}{2}\sigma) \right)' - \frac{\lambda^2}{4 x^2} \bigg[\left(
\ln Q (\ft{i}{2}\sigma) \right)'' + x \left( \ln Q (\ft{i}{2}\sigma) \right)'''
\bigg] + \mathcal{O}(\lambda^3) \right) \, .
\ee
To lowest order in $\lambda$ one has $x(u)=u + \mathcal{O}(\lambda)$ with
$\Delta_\pm^{\rm (ren)}(x) = 1 + \mathcal{O}(\lambda)$ leading to $\Delta_\pm(u)
= (u\pm\ft{i}2)^L$. It is straightforward to verify that the roots of $Q(u)$
verify the modified Bethe equations \re{Bethe-x} to three-loop accuracy \cite{Sta04}.
The apparently unusual feature of \re{ThreeLoopBaxterEq} and \re{Delta-dressed}
compared to \re{Bax0} is that the dressing factors $\Delta_+(u)$ and
$\Delta_-(u)$ depend on derivatives of the function $Q(u)$ evaluated at
$u=\ft{i}2$ and $u=-\ft{i}2$, respectively. We shall elucidate the origin of this
property in a moment.

The auxiliary transfer matrix $t_L(u)$ entering the right-hand side of
\re{ThreeLoopBaxterEq} is a generating function for the conserved charges
\re{q-charges}
\be\label{t1}
t_L (x) = \sqrt{\Delta_+(x)\Delta_-(x)}\left( 2 + \sum_{n \ge 2} q_n (\lambda) \,
x^{- n} \right)
\, .
\ee
Its perturbative expansion starts with \re{t} and includes higher-loop
perturbative corrections to the integrals of motion. In distinction with \re{t},
the series in the right-hand side of \re{t1} does not truncate, thus, reflecting
the asymptotic character of the Baxter equation \re{ThreeLoopBaxterEq}. Notice
the series in the right-hand side of \re{t1} does not involve $\sim x^{-1}$ term.
To see this, it suffices to substitute \re{Q-dressed} and \re{t1} into
\re{ThreeLoopBaxterEq}, expand its both sides at large $u$ and match the
coefficients in front of powers of $u$. In this manner, one deduces from Eqs.\
\re{ThreeLoopBaxterEq} that the `lowest' charge $q_2(\lambda)$ equals\\[-3mm]
\be\label{q2-ren}
q_2 (\lambda)
= -
\left( N + \ft12 L + \ft12 \gamma (\lambda) \right)
\left( N + \ft12 L + \ft12 \gamma (\lambda) - 1 \right)
- \ft14 L
\, ,
\ee\\[-3mm]
where $\gamma (\lambda)$ is given by the sum of two functions
$\gamma_\sigma(\lambda)$ (with $\sigma=\pm$) parameterizing leading asymptotic
behaviour of $\Delta_\sigma^{\rm (ren)}(x)$ at large $x$
\be\label{sum_gamma}
\gamma (\lambda) = \gamma_+(\lambda)-\gamma_-(\lambda)\,,\qquad
 \Delta_\sigma^{\rm (ren)}(x)= \exp\lr{i\gamma_\sigma(\lambda) \,x^{-1} + \mathcal{O}(x^{-2})
 }\,.
\ee
They satisfy the relations $\gamma_+(\lambda)=-\lr{\gamma_-(\lambda)}^*$ and
$\gamma_\sigma(\lambda)=\mathcal{O}(\lambda)$ which ensure that $\gamma
(\lambda)$ takes real values and vanishes for $\lambda\to 0$. As a result, the
charge $q_2 (\lambda)$, Eq.~\re{q2-ren}, also takes real values and approaches
its lowest order value \re{q20} for $\lambda=0$. The reason we used in
\re{sum_gamma} the same notation as in \re{Bethe1} is that $\gamma (\lambda)$
defines the multi-loop anomalous dimension. Combining together \re{Delta-dressed}
and \re{sum_gamma} we get
\be
\label{energy-2} \gamma (\lambda) = \left[ \lambda\frac{d}{du} +
\frac{\lambda^2}{4} \frac{d^3}{du^3} + \frac{\lambda^3}{48} \frac{d^5}{du^5}+
\ldots \right] \big(i\ln Q(u)\big) \bigg|_{\, u=-{i}/2}^{\, u={i}/2} =
\lambda\varepsilon^{(0)} + \lambda^2 \varepsilon^{(1)} + \lambda^3
\varepsilon^{(2)} + \dots \, .
\ee
To leading order in $\lambda$, it coincides with \re{OneLoopEnergy} and matches
\re{Bethe1} up to three loops. One can apply Eqs.~\re{ThreeLoopBaxterEq} and
\re{energy-2} to determine the spectrum of anomalous dimensions of scalar
operators in $\mathcal{N}=4$ SYM theory, Eq.~\re{Def-OpeLoc}, of arbitrary length
$L\ge 2$ and total number of derivatives $N \ge 0$.

Solving the Baxter equation \re{ThreeLoopBaxterEq}, we can determine the function
$Q(u)$ up to three-loop order, Eq.~\re{Q-dressed}, as well as the conserved
charges $q_k (\lambda)$, Eq.~\re{q-charges}. In particular, $q_k(\lambda)$'s take
real values and $Q(u)$ is a real function of $u$. The fact that the leading
function $Q^{(0)} (u)$ has only real roots ensures that the three-loop Bethe
roots $u_k (\lambda)$ are also real. The corresponding quasimomentum is given by
the leading order relation \re{OneLoopEnergy} involving the $Q^{(0)}-$function
and it is protected from perturbative correction in $\lambda$.

To elucidate the physical meaning of the factor $\Delta_\sigma^{\rm (ren)}(x)$,
Eqs.~\re{Delta-dressed} and \re{Delta-ren}, it is instructive to compare the
$SL(2)$ Baxter equation \re{ThreeLoopBaxterEq} with a similar equation describing
the multi-loop anomalous dimension of Wilson operators in the $SU(2)$ sector in
$\mathcal{N}=4$ SYM theory~\cite{BeiDipStau04}, i.e., single-trace operators of
canonical dimension $L$ built from $N$ holomorphic scalars $\mathcal{X} = \phi_3
+ i \phi_4$ and $L - N$ fields $\mathcal{Z} = \phi_1 + i \phi_2$. In that case,
the $SU(2)$ Baxter equation takes the same form as \re{ThreeLoopBaxterEq} with
the only difference that the ``dressing'' factors are merely given by
$\Delta_\pm^{ \scriptscriptstyle \rm SU(2)}(u) = (x(u\mp\ft{i}2))^L$ and do not
involve additional factors similar to $\Delta_\sigma^{\rm (ren)}(x)$. A natural
question arises what is the reason for such difference? We recall that the
anomalous dimensions \re{gamma} are eigenvalues of the dilatation operator which,
in its turn, is one of the generators of the (super)conformal group of the
underlying gauge theory. In the $SU(2)$ sector, the mixing occurs between Wilson
operators carrying the same canonical dimension $L$ and the isotopic charge $N$.
Then, in the multi-color limit, the dilatation operator in this sector can be
mapped into the spin chain with the spin operators being the generators of the
$SU(2)$ subgroup of the full $R$-symmetry group. This should be compared with the
$SL(2)$ sector in which case the dilatation operator is mapped into the spin
chain in such a way that the spin operators are generators of the collinear
$SL(2)$ subgroup and the dilatation operator is one of these generators!

To lowest order in $\lambda$ the recursion works as follows. The {\sl one-loop\/}
dilatation operator is identified as a Hamiltonian of the $SL(2)$ spin chain
with the spin operators being the generators of the collinear subgroup
in gauge theory to {\sl zero-loop\/} order.\footnote{This explains why the
one-loop dilatation operator inherits the conformal symmetry of the classical
Lagrangian. Conformal anomaly affects the anomalous dimensions starting from two
loops only.} To this order, the generators of the collinear subgroup depend on
the classical value of the conformal spin of the scalar field $j=\ft12$ and, as a
consequence, the one-loop dilatation operator coincides with the $SL(2)$ spin
chain of spin $j=\ft12$. Going over to higher orders one expects that the
dilatation operator to $n^{\rm th}$ loop is given by the $SL(2)$ spin chain with
spins corrected by perturbative corrections to the dilatation operator to
$(n-1)^{\rm st}-$loop accuracy. This property finds its manifestation in the
structural form of the Baxter equation \re{ThreeLoopBaxterEq}.

Indeed, the Baxter equation \re{ThreeLoopBaxterEq} involves the dressing factors
$\Delta_\sigma^{\rm (ren)}(x)$, Eq.~\re{Delta-ren}, which depend on the
$Q-$function that satisfies the Baxter equation itself. Replacing $Q(u)$ in
\re{ThreeLoopBaxterEq} by its perturbative expansion \re{Q-dressed} and expanding
$\Delta_\sigma^{\rm (ren)}(x)$ in powers of $\lambda$ it is easy to see that
$\Delta_\sigma^{\rm (ren)}(x)$ induces corrections to the Baxter equation for
$Q^{(n)}(u)$ involving lowest-order functions $Q^{(k)}(u)$ with $0\le k < n$. In
particular, for $n=1$ the one-loop corrections to the conserved charge
$q_2(\lambda)$ depend on the $Q^{(0)}-$function, Eq.\ \re{q2-ren}. We remind that
the charge $q_2^{(0)}$, Eq.~\re{q20}, is related to the total $SL(2)$ conformal
spin $J = N + \ft12 L$ of the scalar operator \re{Def-OpeLoc}. Substituting
$q_2^{(0)}$ in \re{q2-ren} by its explicit expression \re{q20}, one notices that
a part of $\lambda$ correction to $q_2(\lambda)$ proportional to
$\varepsilon^{(0)}$ can be absorbed into the lowest order term as follows
\be\label{q2-ren2}
q_2 (\lambda) = \left[ - (N  + \ft12 L+ \ft{1}{2} \lambda\varepsilon^{(0)}) (N+
\ft12 L + \ft{1}{2} \lambda\varepsilon^{(0)}  - 1) - \ft14 L \right] +
\mathcal{O}(\lambda^2) \, .
\ee
We recall that $\varepsilon^{(0)}$ defines the one-loop correction to the
anomalous dimension of scalar operators, Eq.~\re{gamma}. Going over to higher
orders in $\lambda$, one finds from \re{q2-ren} that $\lambda\varepsilon^{(0)}$
get replaced in \re{q2-ren2} by the multi-loop anomalous dimension \re{energy-2}.
Then, one deduces from \re{q2-ren2} that the factor $\Delta_\sigma^{\rm
(ren)}(x)$ renormalizes the ``bare'' conformal spin $J$ of the scalar operator by
an amount proportional to its anomalous dimension
\be
J=N+\ft12 L\quad  \mapsto \quad J_{\rm ren}=N + \ft 12 L+\ft12 \gamma(\lambda)\,.
\ee
This result can be interpreted as follows. For a conformal operator
$\mathcal{O}_{\rm conf}(0)=\sum_{\bit{\scriptstyle n}} \Psi(\bit{n})
\mathcal{O}_{\bit{\scriptstyle n}} (0)$, its conformal $SL(2)$ spin
$J=\ft12(d+s)$ depends on its scaling dimension, $d$, and projection of its
Lorentz spin on the light-cone, $s$. To order $\lambda^0$ one has $d=N+L$ and
$s=N$ so that $J=N+\ft12 L$. To higher orders in $\lambda$, the spin $s$ is
protected from perturbative corrections while the scaling dimension $d$
receives the anomalous contribution $\gamma(\lambda)$ leading to $J_{\rm ren}
=N + \ft 12 L+\ft12 \gamma(\lambda)$. One can arrive at the same conclusion
from consideration of the conformal Ward identities as explained in details
in Refs.~\cite{BelMul98,BelKorMul04}. In this formalism the additive correction to
the conformal spin proportional to the anomalous dimension comes from the
renormalization of the composite operator given by the product of the Wilson
operators and the trace anomaly of the energy-momentum tensor in regularized
gauge theory~\cite{BelMul98}.

Let us compare \re{energy-2} with the results of explicit diagonalization of the
two-loop mixing matrix in the $\mathcal{N}=4$ SYM, Eqs.~\re{aux} and
\re{mix-mat}. Similar to the Baxter equation, the mixing matrix \re{aux} has
eigenvalues with zero and nonzero quasimomentum. Although the anomalous
dimensions of single trace operators correspond only to the former, we can
perform the comparison for all eigenvalues. In this way, we verified that the two
eigenspectra coincide for $L \le 5$ and $N \le 20$, to two loops at least. Thus,
the relation \re{energy-2} provides the exact solution to the spectral problem
\re{aux} for two-loop mixing matrix in $\mathcal{N}=4$ SYM. Making use of the
relation \re{N2-4}, the correspondence can be further extended to $\mathcal{N}=2$
theory.

We also checked that the ``odd'' conserved charges $q_{2 k + 1}$, corresponding to
the paired eigenvalues have opposite signs in agreement with the lowest order
expectations, Eq.~\re{par}, e.g., for the state with $[L = 3, N = 5]$ given in
Table \ref{ExactSpectra}, one finds for the transfer matrix \re{t1}
\ba
t_{L = 3}^\pm {(u)}
\!\!\!&=&\!\!\!
2 \, u^{3} \pm {u}^{2}
\left( -\ft{1}{12} \, \lambda + \ft{4933}{38016} \lambda^2
\right){\scriptstyle \sqrt{1155}}
\nonumber\\
&+&\!\!\!
u \left( -\ft{73}{2} - \ft{111}{2} \, \lambda
+ \ft{7595}{96} \, \lambda^2 \right)
\pm
( \ft12 + \ft{3791}{1584} \, \lambda - \ft{4894295}{5018112} \, \lambda^2 )
{\scriptstyle \sqrt {1155}}
\, .
\ea
For eigenvalues with zero quasimomentum, we summarized our results in Table
\ref{ExactSpectra}. At two loops, they agree with diagrammatic calculations of
the $[L = 3,N = 2]$ anomalous dimension of Refs.\ \cite{Ede04,EdeJarSok04}, which
is related to the BMN counterpart of the Konishi current \cite{EdeJarSokSta04} due
to multiplet splitting, and the $[L = 3,N = 3]$ result of \cite{Sta04,Ede04,EdeSta06}
as well as with eigenspectra of Ref.\ \cite{Zwi06} based on algebraic construction
of the dilatation operator.

\begin{table}
\renewcommand{\arraystretch}{1.5}
\begin{center}
\begin{tabular}[pos]{||c|c|c|c||}
\hline
\hline
$[L, N]$
&
$\varepsilon^{(0)}$
&
$\varepsilon^{(1)}$
&
$\varepsilon^{(2)}$
\\
\hline \hline $[3, 2]$ & $4$ & $-6 \ {}^{\mbox{\tiny \cite{EdeJarSok04,Ede04}}}$
& $17 \ {}^{\mbox{\tiny \cite{EdeJarSok04}}}$
\\
\hline
$[3, 5]$
&
$\ft{35}{4}$
&
$- \ft{18865}{1152}$
&
$\ft{1068515}{18432}$
\\
\hline
$[3, 8]$
&
$\begin{array}{c} \frac{25}{3} \\ \frac{5087}{420} \end{array}$
&
$\begin{array}{c} - \frac{455}{32} \\ - \frac{1210695307}{49392000} \end{array}$
&
$\begin{array}{c}  \frac{11407175}{248832} \\ \frac{2330723533437143}{26138246400000} \end{array}$
\\
\hline $[4, 2]$
&
$5\pm\sqrt{5}$
&
$ -\ft{1}{2} (17 \pm 5 \sqrt{5})$
&
$\ft{9}{20} (65\pm 23 \sqrt{5})$
\\
\hline
$[4,5]$
&
$\ft{35}{4} \pm \ft{\sqrt{385}}{12}$
&
$- \ft{28139}{1728} \mp \ft{9101 \sqrt{385}}{44352}$
&
$\ft{799837}{13824} \pm \ft{3060649313 \sqrt{385}}{3688312320}$
\\
\hline
$[4,8]$
&
$\begin{array}{l}
6.4113
\\
9.1601
\\
9.9596
\\
9.8710
\\
12.4010
\\
12.9479
\\
12.9650
\\
14.9761
\\
16.4651
\end{array}$
&
$\begin{array}{l}
- 8.4697
\\
- 14.7918
\\
- 18.2198
\\
- 18.6154
\\
- 24.3757
\\
- 25.7258
\\
- 25.2831
\\
- 30.4673
\\
- 33.8137
\end{array}$
&
$\begin{array}{l}
22.4035
\\
47.6639
\\
62.7707
\\
68.0070
\\
88.7702
\\
93.3842
\\
90.0613
\\
111.0666
\\
123.7385
\end{array}$
\\
\hline
$[5, 2]$
&
$\begin{array}{l} 2 \\ 6 \end{array}$
&
$\begin{array}{l} - \ft32 \\ - \ft{21}2 \end{array}$
&
$\begin{array}{l} \ft{37}{16} \\ \ft{555}{16} \end{array}$
\\
\hline \hline
\end{tabular}
\end{center}
\caption{\label{ExactSpectra} Eigenvalues of the one-, two- and three-loop
dilatation operator in $\mathcal{N}=4$ SYM.}
\end{table}

\section{Discussion and conclusions}
\label{Discussion}

Recently, extensive multiloop calculations in various sectors of SYM theories pointed
to persistence of integrability beyond leading perturbative order
\cite{BeiKriSta03,Bei04,EdeJarSok04,Sta04,Ede04,BelKorMul04,EdeJarSokSta04,Zwi06,EdeSta06}.
In this paper, we continued our study of integrability properties of the two-loop
dilatation operator in (supersymmetric) Yang-Mills theories initiated in
Ref.~\cite{BelKorMul04}. As a case of study, we have chosen the sector of
single-trace operators built from holomorphic scalar fields in the
$\mathcal{N}=2$ and $\mathcal{N}=4$ SYM theories and containing an arbitrary
number of covariant derivatives projected onto the light-cone\footnote{The
twist three, spin three anomalous dimension was computed earlier in Refs.\
\cite{Sta04,Ede04,EdeSta06}.}. To one-loop order, in both theories, the dilatation
operator in this sector can be mapped in the multi-color limit into a Hamiltonian of
the $SL(2;\mathbb{R})$ Heisenberg spin chain and its eigenspectrum can be found by means
of the Bethe Ansatz. Our goal was to understand whether integrability survives in high
loops and if so then what are the novel features of the underlying spin chain. To this
end, we performed an explicit two-loop calculation of the dilatation operator and
found that the spectrum of two-loop anomalous dimensions has the same degeneracy
properties as to one loop level. We also demonstrated that, in agreement with our
previous findings \cite{BelKorMul04}, the two-loop anomalous dimensions in
$\mathcal{N}=2$ and $\mathcal{N}=4$ theories differ from each other by an overall
normalization factor indicating that the phenomenon is not sensitive to the
conformal symmetry. These results lay a firm ground to the belief that the
dilatation operator in the two theories is integrable beyond one loop.

As a next step, we tried to uncover integrable structures behind the two-loop
dilatation operator by applying the method of the Baxter $\mathbb{Q}-$operator.
The reason for this is the following. It is well known that in classical
integrable models admitting the Lax representation one can apply the ``magic
recipe''~\cite{Skl90} to perform a canonical transformation to the separated
variables and reduce the original multi-dimensional problem to a set of
one-dimensional ones. In quantum integrable models, the canonical transformation
is replaced by a unitary transformation to the separated coordinates such that
the multi-particle wave function is factorized into a product of single-particle
ones. For the $SL(2;\mathbb{R})$ Heisenberg spin chain (= one-loop dilatation
operator) the representation of the separated coordinates (SoV) has been
constructed in \cite{DerKorMan02}. In this representation, the single-particle wave
function is given by the eigenvalue of the Baxter operator, $Q^{(0)}(u)$, and the
Schr\"odinger equation in the separated variables coincides with the Baxter
equation \re{Bax0} supplemented with \re{Q0} and \re{OneLoopEnergy}. Going over
to higher loops, we assumed that the spin chain describing the eigenspectrum of the
multi-loop dilatation operator admits the SoV representation with the single-particle
wave function $Q(u)$ corrected by perturbative corrections \re{Q-dressed}. In the
$\mathcal{N}=4$ SYM, this is in agreement with the fact that the dilatation operator
for scalar operators with large canonical dimension $L$ and Lorentz spin $N$ can be
identified via the gauge/string correspondence with a Hamiltonian of the classical
sigma-model on AdS${}_3\times$S${}^1$ background. This sigma-model is known to be
completely integrable and it admits both the Lax and SoV representations\footnote{For
an interpretation of the $\mathbb{Q}-$operator in string theory see Ref.~\cite{Gor03}.}
\cite{MikZak78}.

The question remains however how to construct the higher-loop Baxter
$\mathbb{Q}-$operator and what is the analog of the Baxter equation \re{Bax0} for
its eigenvalue $Q(u)$. To answer the second part of this question we first
verified that eigenvalues of two-loop dilatation operator calculated in the
$\mathcal{N}=4$ theory are in agreement with the modified Bethe Ansatz equations
conjectured in \cite{BeiStau05}. Identifying the Bethe roots as roots of $Q(u)$
we worked out a deformed Baxter equation which exactly encodes the one- and
two-loop spectra of anomalous dimension. Then, we demonstrated that the Baxter
equation correctly incorporates a peculiar feature of conformal operators -- the
conformal $SL(2)$ spin of such operators is modified in higher loops by an amount
proportional to their anomalous dimension. From the point of view of spin chains
this property implies that the underlying integrable model is rather unusual --
the Hamiltonian of the spin chain depends on the total $SL(2)$ spin which in its
turn is proportional to the Hamiltonian. Still, to identify this spin chain one
needs the explicit form of the $\mathbb{Q}-$operator. To one-loop order, this
operator has been constructed in \cite{Der99}. Acting on the Wilson operators in
the momentum representation $\widetilde{\mathbb{O}} (\bit{u})$, Eq.~\re{O2}, the
one-loop Baxter operator can be realized as an integral operator
$\mathbb{Q}^{(0)}_u (\bit{u} | \bit{v})$ acting on the momentum fraction, in a
close analogy with the dilatation operator $\mathbb{V}(\bit{u} | \bit{v})$,
Eq.~~\re{measure}. The one-loop evolution kernel $\mathbb{V}^{(0)} (\bit{u} |
\bit{v})$, \re{LO}, arises as a coefficient in the expansion of the kernel
$\mathbb{Q}^{(0)}_u (\bit{u} | \bit{v})$ in the spectral parameter around
$u=\pm\ft{i}2$, in agreement with \re{OneLoopEnergy}. In a similar manner, one
can translate \re{energy-2} into the relation between the two-loop evolution
kernel \re{PerturbExpansionKernel} and the two-loop Baxter operator. Simplicity
of the two-loop kernel \re{V1} gives us a hope that such operator can be
constructed explicitly and the problem deserves additional studies.

\vspace{0.5cm}

\noindent This work was supported by the U.S.\ National Science Foundation under
grant no.\ PHY-0456520 (A.B. and D.M.) and by the Agence Nationale de la Recherche
under grant ANR-06-BLAN-0142-02 (G.K.).

\appendix
\setcounter{section}{0} \setcounter{equation}{0}
\renewcommand{\theequation}{A.\arabic{equation}}

\section*{Appendix}

To one-loop order, the moments of the two-particle kernel
$\mathbb{V}^{(0)}_{12}$, Eq.~\re{plus}, are given by
\be
\Lambda^{(0) m_1 m_2}_{\phantom{(0)} n_1 n_2} = - S_{n_1}^{(1)}
\delta_{n_1}^{m_1} \delta_{n_2}^{m_2} + \frac{n_1 ! n_2!}{m_1 ! m_2!}
\frac{\theta_{n_2 m_2}}{n_2 - m_2} \delta^{m_1 + m_2}_{n_1 + n_2} + \left\{ n_1
\leftrightarrow n_2 \atop m_1 \leftrightarrow m_2 \right\} \, ,
\ee
with discrete step-function taking values $\theta_{n m} = 1$ for $n > m$ and
vanishing otherwise. Analogously, to two loops, the moments of the two-particle
irreducible kernel \re{Res-12-ScaSec-NLO} are given by
\ba
\Lambda^{(1) m_1 m_2}_{\phantom{(1)} n_1 n_2} \!\!\!&=&\!\!\! \left[
S_{n_1}^{(1)} S_{n_1}^{(2)} + S_{n_1}^{(3)} - 2 (4 - \mathcal{N}) S_{n_1}^{(1)}
\right] \delta^{m_1}_{n_1} \delta^{m_2}_{n_2}
+\frac{n_1! n_2!}{m_1! m_2!} \frac{\theta_{n_2 m_2} \delta^{m_1 + m_2}_{n_1 +
n_2}}{n_2 - m_2} \bigg\{ 2 (4 - \mathcal{N})
\\
&-&\!\!\! \frac{1}{2} S^{(2)}_{n_1}-\frac{3}{2} S_{m_1}^{(2)}- \frac{1}{2} \left(
S^{(1)}_{m_1} - S^{(1)}_{n_1} \right) \left( S^{(1)}_{n_1} + 3 S^{(1)}_{m_1} - 4
S^{(1)}_{n_2 - m_2 - 1} \right) \bigg\} + \left\{n_1 \leftrightarrow n_2 \atop
m_1 \leftrightarrow m_2 \right\} \, , \nonumber
\ea
Here the notation was introduced for harmonic sums
\be
S_n^{(1)} = \sum_{\ell=1}^n \frac1\ell=\psi(n+1)-\psi(1)\,,\qquad S_n^{(2)} =
\sum_{\ell=1}^n \frac1{\ell^2}=-\psi'(n+1)+\frac{\pi^2}{6} \,.
\ee
with $\psi(x)=d\ln \Gamma(x)/dx$ being the Euler digamma function.

The moments of the three-particle irreducible kernel $\Lambda^{(1) {m_1 m_2
m_3}}_{\phantom{(1)} {n_1 n_2 n_3}}$ read for $n_2 = 0$
\ba
\Lambda_{\phantom{(1)} {n_1 0 n_3}}^{(1) {m_1 m_2 m_3}} \!\!\!&=&\!\!\!
\frac{n_1! n_3!}{m_1! m_2! m_3! } \frac{ \delta^{m_1 + m_2 + m_3}_{n_1 + n_2 +
n_3} \theta_{n_3 m_3} }{ (n_1 - m_1)(n_3 - m_3) } \Bigg\{ \theta_{m_1 n_1} \left(
\frac{1}{n_1 - m_1} + S^{(1)}_{m_1} - S^{(1)}_{n_1} - S^{(1)}_{m_1 - n_1} \right)
\nonumber\\
&+&\!\!\! \theta_{n_1 m_1} \left( S^{(1)}_{n_1 - m_1} + S^{(1)}_{n_3 - m_3} -
S^{(1)}_{m_2} \right) \Bigg\} + \left\{n_1 \leftrightarrow n_3 \atop m_1
\leftrightarrow m_3 \right\} \, ,
\ea
while for $n_2> 0$ they can be expressed in terms of the $n_2=0$ moments as
\be
\Lambda^{(1) {m_1 m_2 m_3}}_{\phantom{(1)} {n_1 n_2 n_3}} = \sum_{j_1 = 0}^{n_2}
\sum_{j_3 = 0}^{n_2 - j_1} \sum_{k_1 = 0}^{n_2 - j} \sum_{k_3 = 0}^{n_2 - j -
k_1} \frac{ (-1)^{j} n_2! }{j_1! j_3! k_1! k_3! (n_2-j-k)!    }
\Lambda_{\phantom{(1)} {n_1+j_1,\, 0, n_3+j_3}}^{(1){m_1 - k_1, m_2-n_2+j+k, m_3
- k_3}}\,,
\ee
where we used shorthand notations $j=j_1 + j_3$ and $k=k_1 + k_3$.



\end{document}